# Hydrodynamics, particle relabelling and relativity

Peter Holland

Green Templeton College
University of Oxford
Oxford OX2 6HG
England

peter.holland@gtc.ox.ac.uk

**Abstract**

Using the wave equation as an example, it is shown how to extend the hydrodynamic Lagrangian-picture method of constructing field evolution using a continuum of trajectories to second-order theories. The wave equation is represented through Eulerian-picture models that are distinguished by their Lorentz transformation properties. Introducing the idea of the relativity of the particle label, it is demonstrated how the corresponding trajectory models are compatible with the relativity principle. It is also shown how the Eulerian variational formulation may be obtained by canonical transformation from the Lagrangian picture, and how symmetries in the Lagrangian picture may be used to generate Eulerian conserved charges.

PACS: 03.50.-z, 03.65.Ta, 47.10.-g

## 1. Introduction

The hydrodynamic analogy has been instrumental in suggesting new trajectory-based techniques to generate the time-dependence of fields, in particular the quantal wavefunction [1,2]. In the analogy, a field theory that exhibits suitable structure, including a continuity equation (e.g., wave mechanics), is recast as the Eulerian picture of a hydrodynamic model. The aim is to develop a corresponding Lagrangian picture of the fluid and use standard techniques to generate a general solution to the dynamical equations in terms of a continuum of particle trajectories. The word 'particle' is used here in the sense of 'particle-like' – what moves along the trajectories depends on the context and the quantities that are conserved. For example, the trajectories may support mass transfer, or be energy flow lines. In the quantum case, it has been established that a congruence of trajectories – computed independently of the wavefunction (only the initial wavefunction is needed) – may exhibit sufficient structure to provide an exact method to deduce the time-dependence of the wavefunction (including the many-body case) [2].

So far this technique has been carried through for first-order (in time) field theories. The extension to second-order theories, the subject of the present enquiry, presents a problem not encountered previously: how to map the field equations into the first-order

Eulerian hydrodynamic equations. We shall treat the simplest such field-theoretic system: the real one-dimensional wave equation,

$$\frac{\partial^2 \phi}{\partial t^2} - c^2 \frac{\partial^2 \phi}{\partial x^2} = 0, \qquad (1.1)$$

where $c$ is a constant. The Cauchy data will be denoted $\phi_0(x)$ and $\theta(x) = \partial \phi / \partial t \big|_{t=0}$ with boundary conditions $\phi_0, \theta \to 0$ as $x \to \pm\infty$. Our investigation applies whatever the context but for definiteness we suppose $c$ is the speed of light and $\phi(x,t)$ is a Lorentz scalar function.

Our strategy is to write down a Lagrangian-coordinate model whose corresponding Eulerian formulation implies (1.1) when a suitable identification of functions is made. In effect, we replace (1.1) by two coupled first-order Eulerian equations, a continuity equation and a version of Euler's force law that involves only the density and velocity. The latter functions may be connected to $\phi$ in a variety of ways; we focus on three models distinguished by their symmetry properties with respect to Lorentz transformations. Using the Cauchy data in the initial conditions for the Eulerian equations, this provides three distinct formulas to construct $\phi$ from the paths (one of these methods gives $\phi$ up to a sign). Of course, one would hardly adopt this method in practice since the solution to (1.1) may be written down immediately in terms of d'Alembert's formula:

$$\phi(x,t) = \frac{1}{2}\left( \phi_0(x-ct) + \phi_0(x+ct) + \frac{1}{c}\int_{x-ct}^{x+ct} \theta(\tau) d\tau \right). \qquad (1.2)$$

Nevertheless, the method gives a clue as to how higher-dimensional systems may be treated. And our analysis is beneficial for another reason: it enables us to develop a solution, in an elementary case, to an open problem: how the Lagrangian flow is compatible with the relativistic covariance of the Eulerian-picture equations. In discussing this problem in the context of the Lagrangian-coordinate model that underpins Maxwell's equations (regarded as a system of Eulerian equations), it was stated that the Lorentz symmetry arises as a collective effect of the trajectories [3]. This is intuitively reasonable but the nature of the symmetry group of the Lagrangian picture that generates the Eulerian symmetry was left unspecified. We shall show here that the key to resolving this problem is to introduce the *relativity of the particle label*. Then *the Lorentz transformation of the Eulerian picture corresponds to a product of field-dependent and particle label transformations in the Lagrangian picture*. Hence, both pictures are compatible with the relativity principle but through conceptually distinct mechanisms. This also enables us to provide an initial answer to an objection that has been raised against the possibility of defining covariant trajectories associated with an energy-momentum tensor [4]. Developing the approach, we show how the function $\phi$ obeying (1.1) plays a key role in the derivation of the Eulerian canonical theory from the Lagrangian one via a canonical transformation, and how symmetries in the Lagrangian picture generate Eulerian conserved charges.

## 2. Lagrangian picture

In the Lagrangian picture the structure of a continuum, here treated as a fluid, is described through the functional dependence of a particle's position $q(a,t)$ on the time $t$ and on its distinguishing Lagrangian coordinate, or label, $a$ (identified with the position at $t = 0$: $q(a, t = 0) = a$). We assume that the mapping between the two coordinates is single-valued and differentiable with respect to $a$ and $t$. The inverse mapping $a(q,t)$ exists if the deformation gradient (Jacobian) $J = \partial q/\partial a > 0$. The local internal forces due to particle interactions depend on $J$ and its derivatives and each particle responds to a force determined by Newton's second law. Since we shall not restrict the possible applications, we do not worry about the units of the quantities we consider.

Let $\rho_0(a)$ be the initial density of some continuously distributed quantity so that the quantity in an elementary volume $da$ attached to the point $a$ is given by $\rho_0(a)da$. The conservation of this quantity in the course of the motion of the fluid element is expressed through the relation

$$\rho(a,t) = J^{-1}(a,t)\rho_0(a). \tag{2.1}$$

We assume $\rho_0(a) \to 0$ as $a \to \pm\infty$. The prescribed initial data $\rho_0(a)$ and $\partial q(a)/\partial t\big|_{t=0}$ will be adapted to the Cauchy data for (1.1) later on.

To orient our approach, we first consider a one-dimensional relativistic fluid governed by the scalar action $I = \int L\, dt$ where

$$L = \int l(q, \partial q/\partial t, \partial q/\partial a, t)\, da = -\tfrac{1}{2} \int \rho_0(a)\left[c^2 - \left(\frac{\partial q}{\partial t}\right)^2\right]^{1/2} da. \tag{2.2}$$

This describes a free flow; the Euler-Lagrange equation

$$\frac{\partial}{\partial t}\frac{\partial l}{\partial(\partial q/\partial t)} = \frac{\partial l}{\partial q} - \frac{\partial}{\partial a}\frac{\partial l}{\partial(\partial q/\partial a)} \tag{2.3}$$

implies

$$c\rho_0 \frac{\partial}{\partial t}\left[\left(c^2 - \left(\frac{\partial q}{\partial t}\right)^2\right)^{-1/2} \frac{\partial q}{\partial t}\right] = 0 \quad \text{or} \quad \frac{\partial^2 q}{\partial t^2} = 0. \tag{2.4}$$

Each particle $a$ pursues a uniform path. We now include local interactions in the fluid. As we shall see, the following Lagrangian is suitable for generating an Eulerian wave model:

$$L = -\tfrac{1}{2} \int \rho_0^2 J^{-1}\left(c^2 - \left(\frac{\partial q}{\partial t}\right)^2\right) da. \qquad (2.5)$$

The Euler-Lagrange equation (2.3) implies the following density- and velocity-dependent force law:

$$\rho_0 \frac{\partial^2 q}{\partial t^2} = -\frac{\partial}{\partial a}\left(\rho_0 J^{-1}\left(c^2 - \left(\frac{\partial q}{\partial t}\right)^2\right)\right). \qquad (2.6)$$

Although it has a non-relativistic form we shall see later that the action is Lorentz invariant and hence (2.6) is relativistically covariant.

This law has the following properties. Suppose that for each particle the initial density is non-negative and finite and the initial speed is not superluminal:

$$0 \le \rho_0(a) < \infty, \quad \left|\partial q/\partial t\right|_{t=0} \le c \quad \text{for all } a. \qquad (2.7)$$

Then these conditions are preserved for all $t$. This may be proved by observing that (2.6) implies the following two relations:

$$\left(\frac{\partial}{\partial t} + \left(c - \frac{\partial q}{\partial t}\right) J^{-1} \frac{\partial}{\partial a}\right)\left(\rho_0 J^{-1}\left(c + \frac{\partial q}{\partial t}\right)\right) = 0$$

$$\left(\frac{\partial}{\partial t} - \left(c + \frac{\partial q}{\partial t}\right) J^{-1} \frac{\partial}{\partial a}\right)\left(\rho_0 J^{-1}\left(c - \frac{\partial q}{\partial t}\right)\right) = 0. \qquad (2.8)$$

Hence

$$\rho_0 J^{-1}\left(c + \frac{\partial q}{\partial t}\right) = f(q - ct), \quad \rho_0 J^{-1}\left(c - \frac{\partial q}{\partial t}\right) = g(q + ct), \qquad (2.9)$$

for arbitrary functions $f$ and $g$. The initial conditions (2.7) determine the functional forms of $f$ and $g$ and imply that $0 \le f(a), g(a) < \infty$ for all $a$. Adding the two equations (2.9), we obtain

$$2c\rho_0 J^{-1} = f(q - ct) + g(q + ct) \qquad (2.10)$$

which proves that $0 \le \rho < \infty$. Using this result in (2.9) then gives $\left|\partial q/\partial t\right| \le c$ for all $a, t$. Note that, unlike the free case, these results are a collective effect and do not generally follow if (2.7) holds only for a subset of $a$s. Initial global non-superluminal motion is also preserved if $-\infty < \rho_0(a) \le 0$ in (2.7).

## 3. Eulerian picture

Eq. (2.6) is justified by the equations it implies in the Eulerian picture. The density and velocity are obtained from the following formulas:

$$\rho(x,t) = J^{-1}\Big|_{a(x,t)} \rho_0(a(x,t)) \qquad (3.1)$$

$$v(x,t) = \frac{\partial q(a,t)}{\partial t}\bigg|_{a(x,t)}. \qquad (3.2)$$

Differentiating (3.1) and (3.2) with respect to $t$ and using (2.6) we easily deduce the two equations

$$\frac{\partial \rho}{\partial t} + \frac{\partial}{\partial x}(\rho v) = 0 \qquad (3.3)$$

$$\frac{\partial}{\partial t}(\rho v) + c^2 \frac{\partial \rho}{\partial x} = 0. \qquad (3.4)$$

The first equation is the continuity equation and the second (also a continuity equation, corresponding to density $\rho v$ and velocity $c^2/v$ - see Sec. 8) can be expressed, using the first, as a version of Euler's force law,

$$\rho\left(\frac{\partial v}{\partial t} + v \frac{\partial v}{\partial x}\right) = -\frac{\partial}{\partial x}\left[\rho(c^2 - v^2)\right], \qquad (3.5)$$

corresponding to a density $p(\rho,v) = \rho(c^2 - v^2)$.

Eqs. (3.3) and (3.4) are easily solved. They are equivalent to the two relations

$$\left(\frac{\partial}{\partial t} \pm c \frac{\partial}{\partial x}\right)(\rho c \pm \rho v) = 0. \qquad (3.6)$$

Hence

$$\rho(c+v) = f(x-ct), \quad \rho(c-v) = g(x+ct) \qquad (3.7)$$

where $f$ and $g$ are the functions introduced in (2.9). The general solution of the Eulerian equations is therefore

$$\rho(x,t) = (1/2c)[f(x-ct) + g(x+ct)], \quad v(x,t) = (1/2\rho)[f(x-ct) - g(x+ct)] \qquad (3.8)$$

where $f(x) = \rho_0(x)(c + v_0(x))$ and $g(x) = \rho_0(x)(c - v_0(x))$. The Eulerian counterparts of the results proved at the end of Sec. 2 are proved straightforwardly: the initial conditions $0 \le \rho_0(x) < \infty$ and $|v_0(x)| \le c$ for all $x$ are preserved for all time.

**4. Tensorial representation**

We shall present three distinct ways in which the Eulerian equations generate solutions of the scalar wave equation. This is done by identifying two relevant symmetry groups of the equations and associating the Eulerian functions with three alternative sets of Lorentz tensorial components (scalar, vector and rank 2 tensor).

Eqs. (3.3) and (3.4) are covariant with respect to the following two independent groups of transformations: (a) a Lorentz transformation of the coordinates with respect to which the fields are scalars,

$$x' = \gamma(x - ut), \quad t' = \gamma\left(t - \frac{ux}{c^2}\right), \quad \rho'(x',t') = \rho(x,t), \quad v'(x',t') = v(x,t) \quad (4.1)$$

where $\gamma = (1 - u^2/c^2)^{-1/2}$ and (b) a linear transformation of $(\rho c, \rho v)$ with respect to which the coordinates are invariant,

$$x' = x, \quad t' = t, \quad \rho' = \Gamma\left(\rho - \frac{\Lambda}{c^2}\rho v\right), \quad \rho'v' = \Gamma(\rho v - \Lambda \rho). \quad (4.2)$$

The latter implies

$$v' = \frac{v - \Lambda}{1 - \frac{\Lambda}{c^2}v}. \quad (4.3)$$

Here, $u$, $\Lambda$ and $\Gamma$ are independent real constants with $|u| < c$ and we shall henceforth restrict to the case $\Gamma = (1 - \Lambda^2/c^2)^{-1/2}$, $|\Lambda| < c$. The substitution (4.2) is somewhat analogous to a duality transformation (a symmetry of Maxwell's equations involving a linear transformation of the electric and magnetic fields) and will be referred to as such. The product of these groups characterizes the Eulerian relations (3.3) and (3.4) as a Lorentz covariant system implying the wave equation in the following three cases:

*(i) Scalar representation*

In the case $\Lambda = 0$ (so $\Gamma = 1$) the quantities $\rho$ and $\rho v$ are Lorentz scalar fields. It is evident from (3.8) that each of these functions satisfies (1.1) and so either may play the role of $\phi(x,t)$; for definiteness we choose $\phi = \rho$. The initial conditions consistent with the Cauchy data are $\rho_0 = \phi_0$ and, using (3.3), $\rho_0 v_0 = -\int \theta(x) dx$.

*(ii) Vector (current) representation*

In the case $\Lambda = u$ (with $\Gamma = \gamma$) the quantity $J^\mu$, $\mu = 0,1$, with $J^0 = \rho c$, $J^1 = \rho v$, is a Lorentz vector field. With these identifications and writing $x^\mu = (ct, x)$, the Eulerian relations (3.3) and (3.4) state that $J^\mu$ is solenoidal and irrotational in spacetime:

$$\partial_\mu J^\mu = 0 \tag{4.4}$$

$$\varepsilon^{\mu\nu} \partial_\mu J_\nu = 0 \tag{4.5}$$

where $\varepsilon^{\mu\nu} = -\varepsilon^{\nu\mu}$ with $\varepsilon^{01} = 1$, $J_\mu = g_{\mu\nu} J^\nu$, and

$$g_{\mu\nu} = g^{\mu\nu} = \begin{pmatrix} 1 & 0 \\ 0 & -1 \end{pmatrix}. \tag{4.6}$$

Eq. (4.5) gives $\partial_1 J_0 - \partial_0 J_1 = 0$ which implies that there exists a scalar function $\phi(x,t)$ such that $J_\mu = \partial_\mu \phi$. Substituting this result in (4.4) we deduce that $\phi$ obeys the wave equation (1.1). The initial conditions are $\rho_0 c^2 = \theta$, $\rho_0 v_0 = -\partial \phi_0 / \partial x$.

*(iii) Tensor (energy-momentum) representation*

In the scalar and vector models the density has no particular sign and the velocity is not restricted. We now give a physically more intuitive covariant model in which the density is non-negative and the velocity is not superluminal. This method determines a solution of the wave equation up to a sign. Choose $\Lambda = 2u(1 + u^2/c^2)^{-1/2}$ in (4.2). Then the quantity $T^{\mu\nu}$, $\mu, \nu = 0,1$, with $T^{00} = T^{11} = \rho c$, $T^{01} = T^{10} = \rho v$, is a rank 2 symmetric tensor field (note that the equalities of the tensor components are covariant conditions). Together, (3.3) and (3.4) take the form of an energy-momentum continuity equation:

$$\partial_\mu T^{\mu\nu} = 0, \quad \mu, \nu = 0,1. \tag{4.7}$$

To obtain the scalar wave equation from this construction we assume that $\rho \geq 0$ and $|v| \leq c$ (as shown previously, it is enough to assume these conditions initially) and pass to the following new Eulerian variables which are thereby guaranteed to be real:

$$A_0 = (1/\sqrt{2})\rho^{1/2}\left[(c-v)^{1/2} + (c+v)^{1/2}\right]$$
$$A_1 = (1/\sqrt{2})\rho^{1/2}\left[(c-v)^{1/2} - (c+v)^{1/2}\right].$$
(4.8)

The Eulerian relations (3.3) and (3.4) then become

$$\frac{\partial}{c\partial t}\tfrac{1}{2}(A_0^2 + A_1^2) - \frac{\partial}{\partial x}(A_0 A_1) = 0$$
$$-\frac{\partial}{c\partial t}(A_0 A_1) + \frac{\partial}{\partial x}\tfrac{1}{2}(A_0^2 + A_1^2) = 0.$$
(4.9)

Comparing with (4.7) shows that the tensor may be written

$$T_{\mu\nu} = A_\mu A_\nu - \tfrac{1}{2}g_{\mu\nu}A_\sigma A^\sigma \qquad (4.10)$$

with $A^\mu, \mu = 0,1,$ a Lorentz vector. To deduce the scalar wave equation, we insert (4.10) into (4.7) to get

$$A_\nu \partial^\mu A_\mu + A^\sigma (\partial_\sigma A_\nu - \partial_\nu A_\sigma) = 0. \qquad (4.11)$$

Multiplying (4.11) by $A^\nu$ and assuming $A^\nu A_\nu \neq 0$ we obtain $\partial^\mu A_\mu = 0$. Substituting this result into (4.11) gives $\partial_0 A_1 - \partial_1 A_0 = 0$ and hence there exists a scalar function $\phi(x,t)$ such that $A_\mu = \partial_\mu \phi$. From $\partial^\mu A_\mu = 0$ we deduce that $\phi$ obeys (1.1). The quantity (4.10) is therefore the energy-momentum tensor for the massless scalar field. Note that the relations (4.9) would be unaffected if a minus were inserted on the right-hand sides in (4.8) and so the Eulerian variables determine $A^\mu$, and hence $\phi$, only up to a sign. The initial conditions are $\rho_0 c = \tfrac{1}{2}\left[(\partial\phi_0/\partial x)^2 + c^{-2}\theta^2\right]$, $\rho_0 v_0 c = -(\partial\phi_0/\partial x)\theta$.

## 5. Construction of the Eulerian wave from the Lagrangian paths

We show now how the congruence of trajectories obeying (2.6) may be employed to construct the time-dependence of the scalar wave. In this method, the trajectories are first used to determine the Eulerian density and velocity, (3.1) and (3.2). From these functions, we may deduce the wave in any of the three ways described in the last section. In each of the three cases the sets of trajectories obey the same law of motion but differ due to the variable initial conditions. These different ways of modelling the scalar wave equation will have correspondingly distinct interpretations; in the tensor case the paths may be regarded as energy flow lines of the scalar field, for example. Writing $\dot{q} = \partial q/\partial t$, the scalar, vector and tensor representations give, respectively,

$$\phi(x,t) = \left(J^{-1}\rho_0\right)_{a(x,t)} \qquad (5.1)$$

where $\rho_0 = \phi_0$ and $\rho_0 v_0 = -\int \theta(a)da$,

$$\phi(x,t) = \int J_0 dt + J_1 dx = \int \left(c^2 J^{-1}\rho_0\right)_{a(x,t)} dt - \left(\dot{q}J^{-1}\rho_0\right)_{a(x,t)} dx \tag{5.2}$$

where $\rho_0(a)c^2 = \theta$ and $\rho_0 \dot{q}_0 = -\partial\phi_0/\partial a$, and

$$\pm\phi(x,t) = \left(1/\sqrt{2}\right)\int \left\{c\left(J^{-1}\rho_0\right)^{1/2}\left[(c-\dot{q})^{1/2} + (c+\dot{q})^{1/2}\right]\right\}_{a(x,t)} dt$$
$$+ \left\{\left(J^{-1}\rho_0\right)^{1/2}\left[(c-\dot{q})^{1/2} - (c+\dot{q})^{1/2}\right]\right\}_{a(x,t)} dx \tag{5.3}$$

where $\rho_0 c = \tfrac{1}{2}\left[(\partial\phi_0/\partial a)^2 + c^{-2}\theta^2\right]$ and $\rho_0 v_0 c = -(\partial\phi_0/\partial a)\theta$.

To do the computations we note that, rewriting (2.9), the trajectories may be found by solving the following coupled first-order equations:

$$\frac{\partial q}{\partial a} = \frac{2c\rho_0}{f(q-ct) + g(q+ct)}, \quad \frac{\partial q}{\partial t} = \frac{c[f(q-ct) - g(q+ct)]}{f(q-ct) + g(q+ct)} \tag{5.4}$$

where $f$ and $g$ are fixed by the Cauchy conditions via

$$f(a) = \rho_0\left(c + \frac{\partial q_0}{\partial t}\right), \quad g(a) = \rho_0\left(c - \frac{\partial q_0}{\partial t}\right). \tag{5.5}$$

Actually, in this simple case the quantities $J^{-1}\rho_0$ and $\dot{q}$ appearing in the expressions (5.1)-(5.3) are given in terms of the known functions $f$ and $g$ by (5.4) so we do not need the explicit solution $q(a,t)$. Inserting in turn in (5.5) the three sets of initial conditions, and substituting the resulting expressions for $f$ and $g$ into (5.4), it is easy to deduce the expression (1.2) for each of the functions $\phi$.

### 6. Implementing the relativity principle in the Lagrangian picture

In the Lagrangian picture the relativity principle is implemented by a transformation of the independent variables $a$ and $t$ and their dependent functions. We will show that this procedure corresponds to a Lorentz transformation of the independent variables of the Eulerian picture, $x$ and $t$, and their dependent tensor functions. The transformation functions $t' = t'(a,t)$ and $a' = a'(a,t)$ may depend on $q(a,t)$; they will then be *field-dependent* and generally nonlinear.

An example of the sort of problem that is of concern is as follows. The possibility of deriving a Lorentz covariant system of Eulerian equations from the trajectory model, and in particular the formulas of the last section, raises the question of how the Lorentzian symmetry emerges from properties of the particle model when the respective pictures are connected. Indeed, we seem to have a paradox. In the rank 2 tensor case, for example, the formula (3.2) connecting the velocities becomes

$$\left.\frac{\partial q(a,t)}{\partial t}\right|_{a(x,t)} = \frac{T^{01}(x,t)}{T^{00}(x,t)}. \tag{6.1}$$

Now suppose we work with the Eulerian equations in a Lorentz-transformed frame where the variables are denoted by a prime. Since there is nothing special about the first Lorentz frame, there exist functions $q'(a',t')$, $\rho'_0(a')$ for which we can repeat the above deduction of the Eulerian theory starting from a primed version of (2.6) and such that

$$\left.\frac{\partial q'(a',t')}{\partial t'}\right|_{a'(x',t')} = \frac{T'^{01}(x',t')}{T'^{00}(x',t')}. \tag{6.2}$$

The relation between the right-hand sides of (6.1) and (6.2) is well defined, following from the tensorial properties of the quantities concerned. How are the trajectory functions $q$ and $q'$ and their derivatives connected? Clearly, (6.1) and (6.2) cannot both be valid if the left-hand sides are connected by the usual transformation rule for a Lorentz vector. We shall show that we may maintain the simultaneous validity of (6.1) and (6.2) by introducing the notion of the relativity of the particle label. The paradox is resolved because the Lorentz transformation of $q$ is accompanied by a suitable transformation of the label $a$.

Formally, the situation is analogous to the covariant formulation of QED in a non-Lorentz covariant gauge (e.g., $\nabla \cdot \mathbf{A} = 0$). There, the gauge condition, the analogue of the equation of motion (6.1), is maintained if when Lorentz transformed the potential undergoes a suitable accompanying gauge transformation, in sum a 'gauge-dependent Lorentz transformation' (that yields $\nabla' \cdot \mathbf{A}' = 0$, the analogue of (6.2)). We shall use a similar terminology here, replacing 'gauge' by 'label'.

Three points are noteworthy in our analysis:

(a) We are able to obtain a range of tensorial representations of the density and velocity in the Eulerian picture because their transformations may be decoupled from those of the independent variables $x$, $t$. In contrast, in the Lagrangian picture the position $q$ is a function of the independent variables $a,t$ and so its transformation induces unique transformation rules for the velocity and deformation $J$ (and hence for $\rho$ once the transformation of $\rho_0(a)$ is specified). These disparate behaviours in the two pictures are reconciled by introducing an additional label transformation in the Lagrangian case that enables us to mirror the freedom to transform the Eulerian functions.

(b) A unique transformation of the Eulerian independent and dependent variables generally corresponds to a *class* of transformations of the Lagrangian independent and dependent variables. This is expected since the Eulerian picture is a reduction of the Lagrangian one [5]. This means that the Eulerian identity does not correspond to the Lagrangian identity but to a relabelling class. Continuous Lagrangian transformations corresponding to Eulerian transformations will have relabelling as a component, independent of the parameters defining a Eulerian transformation.
(c) Relativistic transformation laws for the Lagrangian velocity and deformation gradient can be obtained both by an 'external' transformation (of the current position and the time) and by an 'internal' transformation (of the label) for which the position may be invariant.

We now show how the relativity principle is implemented in the Lagrangian picture by finding the symmetries that correspond to the Eulerian ones described earlier.

(i) *Pure label transformation (analogue of Eulerian identity transformation)*

The identity transformation in the Eulerian picture,

$$x' = x, \quad t' = t, \quad v'(x',t') = v(x,t), \quad \rho'(x',t') = \rho(x,t), \qquad (6.3)$$

corresponds to the following transformation in the Lagrangian picture:

$$a' = a'(a,t), \quad t' = t, \quad q'(a',t') = q(a,t), \quad \frac{\partial q'}{\partial t'} = \frac{\partial q}{\partial t}, \quad \rho'_0(a')\left(\frac{\partial q'}{\partial a'}\right)^{-1} = \rho_0(a)\left(\frac{\partial q}{\partial a}\right)^{-1}. \qquad (6.4)$$

To discover the function $a'$, note that the first three members of (6.4) imply

$$\frac{\partial q'}{\partial a'} = \left(\frac{\partial a'}{\partial a}\right)^{-1}\frac{\partial q}{\partial a} \qquad (6.5)$$

$$\frac{\partial q'}{\partial t'} = \frac{\partial q}{\partial t} - \left(\frac{\partial a'}{\partial a}\right)^{-1}\frac{\partial a'}{\partial t}\frac{\partial q}{\partial a}. \qquad (6.6)$$

Comparing with the last two members of (6.4) we obtain

$$\rho'_0(a')\frac{\partial a'}{\partial a} = \rho_0(a), \quad \frac{\partial a'}{\partial t} = 0. \qquad (6.7)$$

Thus, the corresponding transformation in the Lagrangian picture constitutes a time-independent diffeomorphism $a'(a)$, or relabelling of the fluid particles, with respect to which the reference density transforms as a tensor density. Invariants of the transformation include the associated quantity in an elementary volume of label space,

$$\rho'_0(a')da' = \rho_0(a)da, \qquad (6.8)$$

and the position and velocity.
  The choice of the initial particle position as label, $q_0(a) = a$, is intuitively reasonable and implies that the element $da$ coincides with a spatial volume. Since, according to (6.4), $q'(a',t'=0) = q(a,t=0)$ and, in general, $a \neq a'$, the transformation (6.7) expresses the freedom to choose a label other than the initial position. We can use this symmetry to simplify the Lagrangian-picture problem without affecting the Eulerian formulation (as we shall see below, time-dependent label transforms do not generally leave the Eulerian functions invariant). For example, we can transform to a labelling

$$a' = \int_{-\infty}^{a} \rho_0(a) da \qquad (6.9)$$

so that the density is uniform: $\rho_0'(a') = 1$. Note that, starting from an arbitrary labelling, we can always transform to a labelling which coincides with initial position: in general, $q(a,t=0) = f(a)$ so we just set $a' = f(a)$.

A general transformation $a'(a)$ clearly leaves the action $I$, with Lagrangian (2.2) or (2.5), invariant, so the equation of motion (2.6) has a 'generally covariant' form with respect to arbitrary relabellings. A different uniformizing transformation is described below in (iii).

(ii) *Field-dependent Lorentz transformation (analogue of Lorentz transformation)*

The analogue of the Lorentz transformation of the Eulerian independent variables is, assuming no relabelling is involved,

$$a' = a, \quad t' = \gamma\left(t - \frac{uq(a,t)}{c^2}\right), \quad q'(a',t') = \gamma(q(a,t) - ut), \quad \rho_0'(a') = \rho_0(a). \qquad (6.10)$$

These transformations imply

$$\frac{\partial q'}{\partial t'} = \left(\frac{\partial q}{\partial t} - u\right)\left(1 - \frac{u}{c^2}\frac{\partial q}{\partial t}\right)^{-1} \qquad (6.11)$$

$$\rho_0'(a')\left(\frac{\partial q'}{\partial a'}\right)^{-1} = \gamma \rho_0(a)\left(\frac{\partial q}{\partial a}\right)^{-1}\left(1 - \frac{u}{c^2}\frac{\partial q}{\partial t}\right). \qquad (6.12)$$

Hence, when suitably combined, the corresponding Eulerian density and velocity transform as a vector, with $\Lambda = u$ in (4.2). Noting that the transformed integration element is $da'dt' = \gamma\left(1 - (u/c^2)(\partial q/\partial t)\right)dadt$ it is easy to prove the invariance of the action $I$, with Lagrangian (2.2) or (2.5), under this transformation.

We can always make a transformation to a label $a'(a)$ so that the condition $q(a,t=0) = a$ is preserved under the transformation (6.10): From the second member of (6.10) we set $t'(a,t) = 0$ and solve to find the corresponding time $t = f(a,u)$ in the original frame. Choosing $a' = c^2\gamma^{-1}f(a,u)/u$ we then have, from the third member, $q'(a',t'=0) = a'$.

(iii) *Generalized label transformation (analogue of duality transformation)*

Our task now is to find the Lagrangian analogue of the duality transformation (4.2) and (4.3). We therefore seek a transformation

$$a' = a'(a,t), \quad t' = t, \quad q'(a',t') = q(a,t) \tag{6.13}$$

which generates the Lagrangian analogues of (4.2) and (4.3):

$$\rho_0'(a')\left(\frac{\partial q'}{\partial a'}\right)^{-1} = \Gamma \rho_0(a)\left(\frac{\partial q}{\partial a}\right)^{-1}\left(1 - \frac{\Lambda}{c^2}\frac{\partial q}{\partial t}\right) \tag{6.14}$$

$$\frac{\partial q'}{\partial t'} = \left(\frac{\partial q}{\partial t} - \Lambda\right)\left(1 - \frac{\Lambda}{c^2}\frac{\partial q}{\partial t}\right)^{-1}. \tag{6.15}$$

Comparing with (6.5) and (6.6) we obtain

$$\rho_0'(a')\frac{\partial a'}{\partial a} = \Gamma \rho_0(a)\left(1 - \frac{\Lambda}{c^2}\frac{\partial q}{\partial t}\right), \quad \rho_0'(a')\frac{\partial a'}{\partial t} = \Gamma \frac{\Lambda}{c^2}\rho_0(a)J^{-1}\left(c^2 - \left(\frac{\partial q}{\partial t}\right)^2\right). \tag{6.16}$$

These equations both constrain $a'$ and determine the transformation law of $\rho_0(a)$. They generalize the relabelling equations (6.9) to which they reduce when $\Lambda = 0$, and it is readily checked that they leave invariant the action *I*, with Lagrangian (2.2) or (2.5).

Using (2.6) and the boundary conditions, the relations (6.16) may be written

$$\rho_0'(a')\frac{\partial a'}{\partial a} = \frac{\partial K}{\partial a}, \quad \rho_0'(a')\frac{\partial a'}{\partial t} = \frac{\partial K}{\partial t} \tag{6.17}$$

where

$$K(a,t,\Lambda,\Gamma) = \Gamma \int_{-\infty}^{a} \rho_0(a)\left(1 - \frac{\Lambda}{c^2}\frac{\partial q}{\partial t}\right) da. \tag{6.18}$$

A class of transformations is obtained if we eliminate $\rho_0'(a')$ to get

$$\frac{\partial a'}{\partial a}\frac{\partial K}{\partial t} - \frac{\partial a'}{\partial t}\frac{\partial K}{\partial a} = 0. \tag{6.19}$$

The totality of solutions is given by [6]

$$a' = w[K(a,t)] \tag{6.20}$$

where $w$ is an arbitrary function of the function $K$. The transformation is completed by observing from (6.17) that

$$\rho'_0(a') = (dw[K(a,t)]/dK)^{-1}. \tag{6.21}$$

When $w = K$ this transformation uniformizes the reference density, generalizing (6.9).

(iv) *Label-dependent Lorentz transformation*

The Lagrangian-picture transformations that correspond to the three Eulerian-picture tensor transformations described in Sec. 4 may now be stated. They are combinations of (ii) and (iii) above, i.e.,

$$a' = a(a,t,\Lambda), \quad t' = \gamma\left(t - \frac{uq(a,t)}{c^2}\right), \quad q'(a',t') = \gamma(q(a,t) - ut), \tag{6.22}$$

and are distinguished by the choice of parameter $\Lambda$. The corresponding transformation of the Eulerian functions $\rho$ and $v$ comprises a Lorentz transformation (4.1) of their arguments with parameter $u$ and a duality transformation (4.2) with the parameter $\Lambda$ replaced by $(u+\Lambda)(1+u\Lambda/c^2)^{-1}$.

*Scalar*: We use (6.16) to 'undo' the relativistic transformation (6.11) and (6.12). This is achieved by choosing $\Lambda = -u$ (and hence $\Gamma = \gamma$).

*Vector*: Where we choose $\Lambda = u$ and $\Gamma = \gamma$ in the Eulerian case in Sec. 4 (ii), in the Lagrangian case the vector nature of the current is already implied by the $t,q$ transformation so we choose $\Lambda = 0$ (and so $\Gamma = 1$) in (6.16).

*Tensor*: We have to achieve the opposite of the scalar case and use (6.16) to enhance the relativistic transformation of (6.11) and (6.12) to obtain the Eulerian parameter $2u(1+u^2/c^2)^{-1}$. We choose $\Lambda = u$ (and so $\Gamma = \gamma$) in (6.16).

**7. Eulerian variational principle obtained by canonical transformation from the Lagrangian picture**

Passing to Eulerian variables, the Lagrangian (2.5) becomes

$$L = -\tfrac{1}{2}\int \rho^2(c^2 - v^2)dx. \tag{7.1}$$

For each of the tensorial models described in Sec. 4 the integrand is a Lorentz scalar and hence so is the action (the same is true for the free case based on (2.2)). But this action is not suitable for deriving the fluid equations since no time derivatives appear, and $\rho$ and $v$ are not canonical variables, well known problems in Eulerian continuum mechanics. A common way of remedying this is to introduce fluid potentials via Lagrangian multipliers. This results in a satisfactory Eulerian variational formalism but is somewhat *ad hoc*. Following previous work [2,7] we shall show how the Eulerian action principle may be obtained in a natural way through a canonical transformation from the Lagrangian-picture action principle. This procedure indicates that the function $\phi$ obeying (1.1) plays a key role in the Eulerian formalism, for the potential that is the canonical partner to $\rho$ is $-\phi$.

In the Lagrangian picture, the canonical field momenta are

$$p(a) = \frac{\partial L}{\partial(\partial q(a)/\partial t)} = \rho_0(a)^2 J^{-1} \frac{\partial q(a)}{\partial t} \qquad (7.2)$$

and so, following a Legendre transformation, the Hamiltonian is

$$H[q,p,t] = \int p(a) \frac{\partial q(a)}{\partial t} da - L[q, \partial q/\partial t, t]$$
$$= \int \left[ \frac{p(a)^2}{2\rho_0(a)^2 J^{-1}} + \tfrac{1}{2} c^2 \rho_0(a)^2 J^{-1} \right] da. \qquad (7.3)$$

Hamilton's equations are

$$\frac{\partial q(a)}{\partial t} = \frac{\delta H}{\delta p(a)}, \quad \frac{\partial p(a)}{\partial t} = -\frac{\delta H}{\delta q(a)} \qquad (7.4)$$

which when combined reproduce (2.6).

We obtain a corresponding canonical theory in the Eulerian picture via a suitably chosen canonical transformation linking the old phase space coordinates $q(a), p(a)$ to a new set denoted by $Q(x), P(x)$ from which the functions $\rho(x), v(x)$ may be derived. If the generating function of the canonical transformation is a time-independent functional of the old coordinates and the new momenta, $W[q(a), P(x)]$, the transformation formulas are

$$Q(x) = \frac{\delta W}{\delta P(x)}, \quad p(a) = \frac{\delta W}{\delta q(a)}, \quad K[Q,P,t] = H[q,p,t]. \qquad (7.5)$$

A suitable generating function is given by

$$W = \int \delta(x - q(a)) \rho_0(a) P(x) da\, dx. \qquad (7.6)$$

Our transformation equations are then

$$Q(x) = \int \delta(x - q(a)) \rho_0(a)\, da = \left(J^{-1}\rho_0\right)\Big|_{a(x)}, \qquad (7.7)$$

and

$$p(a) = \rho_0(a) \frac{\partial P(x)}{\partial x}\bigg|_{x(a)}. \qquad (7.8)$$

Combining these formulas and substituting in (7.3) the new Hamiltonian is

$$K[Q,P,t] = \frac{1}{2}\int\left[\left(\frac{\partial P(x)}{\partial x}\right)^2 + c^2 Q(x)^2\right]dx. \qquad (7.9)$$

From (7.7) we observe that the coordinate $Q(x)$ may be identified with $\rho(x)$ and using (7.2) and (3.2) the expression (7.8) for the momentum gives for the Eulerian velocity

$$v(x) = \frac{1}{Q(x)}\frac{\partial P(x)}{\partial x}. \qquad (7.10)$$

Hamilton's equations for the new variables are

$$\frac{\partial Q}{\partial t} = \frac{\delta K}{\delta P} = -\frac{\partial^2 P}{\partial x^2}, \quad \frac{\partial P}{\partial t} = -\frac{\delta K}{\delta Q} = -c^2 Q. \qquad (7.11)$$

Using (7.10) to eliminate $P$ we recover the Eulerian equations (3.3) and (3.4). On the other hand, eliminating $Q$, we find that $P$ satisfies (1.1). Hence, we see that the generating function (7.6) produces the vector representation of Sec. 4 (ii), with $P$ identified with the potential $-\phi$. Note that the continuity equation (3.3) comes out as one of Hamilton's equations whereas its Lagrangian-picture counterpart (2.1) does not appear in (7.4). This occurs because of the identification of $Q(x)$ with $\rho(x)$.

An inverse Legendre transformation yields the Lagrangian in the Eulerian picture:

$$L[Q,P,t] = \int P\frac{\partial Q}{\partial t}dx - K[Q,P] = -\int\left[\phi\frac{\partial \rho}{\partial t} + \frac{1}{2}\left(\frac{\partial \phi}{\partial x}\right)^2 + \frac{1}{2}c^2\rho^2\right]dx. \qquad (7.12)$$

where we have replaced $P$ by $-\phi$ and $Q$ by $\rho$. Varying $\rho$ and $\phi$, the Euler-Lagrange equations reproduce Hamilton's equations (7.11). Substituting for $Q = \rho$ from (7.11), $L$ may be brought to the form

$$L = -\frac{1}{2}\int \partial_\mu \phi \partial^\mu \phi \, dx, \quad (7.13)$$

which is the standard Lagrangian for the massless scalar field whose Euler-Lagrange equation is (1.1). It evidently coincides with (7.1).

## 8. Conserved charges

One of the benefits of the Lagrangian-coordinate approach to continuum mechanics is that it provides an additional means to discover Eulerian conserved charges, particularly through the application of Noether's theorem. This is especially useful in cases where the Eulerian-picture transformation corresponding to a Lagrangian-picture symmetry is trivial.

Consider the one-parameter transformation

$$t' = t + \varepsilon \xi_0(a,t), \quad a' = a + \varepsilon \xi(a,t), \quad q'(a',t') = q(a,t) + \varepsilon \eta(a,t) \quad (8.1)$$

where $\varepsilon$ is a dimensionless infinitesimal parameter. Note that the functional dependence of the transformation functions $\xi_0$, $\xi$ and $\eta$ may come from $q(a,t)$ and its derivatives. We also need the infinitesimal transformation of the reference density, which is in general

$$\rho_0'(a') = \rho_0(a) + \varepsilon P(a) + \varepsilon \frac{\partial \rho_0}{\partial a} \xi(a,t) \quad (8.2)$$

where $P(a)$ represents the functional variation: $P(a) = \rho_0'(a) - \rho_0(a)$. In the present context we consider only transformations due to variations in the coordinates, i.e., $\rho_0$ is form invariant, so $P(a) = 0$. The transformation functions are constrained by the requirement that the action is invariant, which implies a set of symmetries of the dynamical equations. Noether's theorem asserts that for each symmetry the charge $\int_{-\infty}^{\infty} C(a,t)\,da$, where the density is given by

$$C(a,t) = \xi_0 l + \left(\eta - \xi_0 \frac{\partial q}{\partial t} - \xi \frac{\partial q}{\partial a}\right) \frac{\partial l}{\partial(\partial q/\partial t)}, \quad (8.3)$$

obeys the conservation law

$$\frac{d}{dt}\int_{-\infty}^{\infty} C(a,t)\,da = 0. \quad (8.4)$$

We can convert this into a corresponding Eulerian conserved density via the formula $E(x,t) = C(a,t) J^{-1}(a,t)\big|_{a(x,t)}$ with

$$\frac{d}{dt}\int_{-\infty}^{\infty} E(x,t)\,dx = 0. \tag{8.5}$$

The latter may be compared with the conserved charge obtained directly in the Eulerian formulation using the symmetry transformation that corresponds to (8.1). With reference to the Lagrangian (7.12), we consider the transformation

$$t' = t + \varepsilon\theta_0(x,t), \quad x' = x + \varepsilon\theta(x,t), \quad \rho'(x',t') = \rho(x,t) + \varepsilon P(x,t),$$
$$\phi(x',t') = \phi(x,t) + \varepsilon\Phi(x,t). \tag{8.6}$$

Denoting the Lagrangian density by $\bar{l}$, the conserved density implied by Noether's theorem is

$$\bar{E}(x,t) = \theta_0 \bar{l} + \left(P - \theta_0 \frac{\partial\rho}{\partial t} - \theta\frac{\partial\rho}{\partial x}\right)\frac{\partial \bar{l}}{\partial(\partial\rho/\partial t)}. \tag{8.7}$$

We shall consider the infinitesimal versions of the symmetries given previously:

(i) *Pure label transformation*: $\xi_0 = \eta = 0, \xi = \xi(a)$. Using (8.2), the infinitesimal form of (6.7) is

$$\frac{\partial}{\partial a}(\rho_0 \xi) = 0 \tag{8.8}$$

whose solution is $\xi = -k/\rho_0$ where $k$ is a constant. Thus $C(a,t) = k\rho_0\, \partial q/\partial t$ and $E(x,t) = k\rho v$ which is (proportional to) the linear momentum density. The continuity equation obeyed by the latter is (3.4). The corresponding Eulerian transformation is the identity with no useful associated conserved quantity.

(ii) *Lorentz transformation*: $\varepsilon = u/c, \xi_0 = -q/c, \xi = 0, \eta = -ct$. Then

$$C(a,t) = \rho_0^2 J^{-1}\left(\tfrac{1}{2}qc + \tfrac{1}{2}\frac{q}{c}\left(\frac{\partial q}{\partial t}\right)^2 - ct\frac{\partial q}{\partial t}\right) \tag{8.9}$$

whence

$$E(x,t) = \rho^2\left(\tfrac{1}{2}xc + \tfrac{1}{2}\frac{x}{c}v^2 - ctv\right). \tag{8.10}$$

The corresponding Eulerian transformation is defined by $\theta_0 = -x/c, \theta = -ct, P = \partial\phi/c\partial x, \Phi = 0$. This gives

$$\bar{E}(x,t) = \frac{x}{c}\left(\frac{1}{2}\left(\frac{\partial\phi}{\partial x}\right)^2 + \frac{1}{2}c^2\rho^2\right) + ct\rho\frac{\partial\phi}{\partial x} - \frac{\partial}{\partial x}\left(\frac{1}{2c}\phi^2 + ct\phi\rho\right) \qquad (8.11)$$

which, noting that $\partial\phi/\partial x = -\rho v$, agrees with (8.10) apart from an additional divergence. The latter expresses the lack of uniqueness of the local Noether current and does not affect the global charge. Note that $E(x,t) = (x/c)T^{00} - tT^{01}$ where $T^{\mu\nu}$ is the tensor (4.10) for the massless field, a component of the relativistic angular momentum.

(iii) *Duality*: $\varepsilon = \Lambda/c, \xi_0 = \eta = 0, \xi = \xi(a,t)$. Then (6.16) gives

$$\frac{\partial}{\partial a}(\rho_0\xi) = -\frac{1}{c}\rho_0\frac{\partial q}{\partial t}, \quad \rho_0\frac{\partial\xi}{\partial t} = \frac{1}{c}\rho_0 J^{-1}\left(c^2 - \left(\frac{\partial q}{\partial t}\right)^2\right). \qquad (8.12)$$

This is satisfied by

$$\rho_0\xi = -\frac{1}{c}\int^a \rho_0\frac{\partial q}{\partial t}da - k \qquad (8.13)$$

so

$$C(a,t) = \frac{1}{2c}\frac{\partial}{\partial a}\left(\int^a \rho_0\left(\frac{\partial q}{\partial t}\right)da\right)^2 + k\rho_0\frac{\partial q}{\partial t} \qquad (8.14)$$

which implies the same global charge as the pure label transformation (i). The Lagrangian (7.12) corresponds to the vector case of Sec. 6 so the corresponding Eulerian transformation is the identity.

## 9. Comments

The results presented here form part of a wider programme examining whether the formal structure of continuum mechanics – which comprises two major pictures, the particle (Lagrangian) and the field (Eulerian) – is mirrored in other field theories that hitherto have been presented only in a language that corresponds to the Eulerian picture. Apart from providing novel ways to solve the Eulerian – or field – equations, a motivation for developing a Lagrangian version of a field theory, if it exists, is to widen the conceptual landscape of that theory. The approach is particularly potent in the quantum case for it suggests and justifies the introduction of the trajectory concept in that discipline.

In the manifestly covariant relativistic Lagrangian description of a (one-dimensional) fluid, the history of a fluid element is represented by the spacetime coordinates $q^\mu(a,\tau), \mu = 0,1$, where the invariants $a$ and $\tau$ are, respectively, the particle label and the evolution parameter [8-11]. This is an obvious extension of the multiple-time description of a discrete many-body system in which each particle is attributed its own time coordinate. With an eye on an eventual application of the theory of this paper to quantum nonlocality, here we attribute a common time $t$ ($\equiv q^0 = \tau$) to all the particles so the process is described just by the function $q(a,t)$ ($\equiv q^1$). Whether a manifestly covariant version can be developed will be the subject of a further study.

We might regard the single-time theory as a continuum analogue of 'predictive mechanics', an approach to relativistic many-body mechanics developed in the 1960s (see the review [12] and references therein) when it was shown that attributing a common time to the particles has a Lorentz covariant content when the interparticle forces obey (the 'Currie-Hill') constraints. These conditions have apparently not been given for a non-denumerable system of particles but the continuous model described here presumably obeys the putative constraints. In our continuum model the action-at-a-distance interaction of the discrete particle system is replaced by a local action extending between each particle and its neighbours, the interparticle force involving a differential in the particle label. In addition, a novel element enters through a continuous symmetry group not available for discrete systems, represented by a tensor calculus associated with the particle label which need not be an invariant.

The particle label transformation is crucial to establishing the covariance of the Lagrangian picture in the rank 2 tensor representation of the velocity. However, the method works in the example we have studied because of a fortuitous property of the one-dimensional theory, namely, $T^{00} = T^{11}$, which ensures that all the tensor components have an interpretation in terms of the basic hydrodynamic variables (density and velocity). Extending the method to higher dimensions will require broadening the hydrodynamic representation in order to interpret all the tensor components.

Finally, we observe that the technique inaugurated here may be pertinent to the outstanding problem of developing relativistically covariant hidden variable models in quantum mechanics. Traditionally, this enquiry has been dominated by the question: 'can hidden variables be made Lorentz covariant?' Following our investigation here, the query might be better posed: 'can hidden variables be made compatible with the relativity principle in a way that is consistent with the Lorentz covariance of quantum theory?'. Label transforms have hitherto been ignored in the most successful trajectory hidden variable theory, that of de Broglie and Bohm, where it has generally been believed that a preferred frame of reference is required.